# Possibility of Land Movement Prediction for Creep or before Earthquake Using Lidar Geodetic Data in a Machine Learning Scheme

M. Kiani

*Abstract*—Earthquake prediction is one of the most pursued problems in geoscience. Different geological and seismological approaches exist for the prediction of the earthquake and its subsequent land change. However, in many cases, they fail in their mission. In this paper, we address the well-established earthquake prediction problem by a novel approach. We use a four-dimensional location-time machine learning scheme to estimate the time of earthquake and its land change. We present a study for the Ridgecrest, California 2019 earthquake prediction. We show the accuracy of our method is around 14 centimeters for the land change, and around 2 days for the time of the earthquake, predicted from data more than 3 years before the earthquake.

## I. INTRODUCTION

EARTHQUAKES and creeps in an area are of fundamental importance, since they, especially the former, affect people's lives. The underlying mechanisms of these naturally occurring phenomena are not fully understood. Hence, their prediction is hampered by a multitude of problems. Some go even further to say earthquakes are unpredictable [1]. However, many attempts have been made on this problem [2], [3]. Main categories of the approaches in earthquake and creep prediction are seismological, geological, and geodetic.

Seismological approaches base their prediction on the preseismic movements and wave-propagation concepts [4], [5], [6]. These approaches tend to fail in their mission in many cases. One such example is the Tohoku earthquake in which the smaller earthquake that happened 2 days before the main 9 Mw shock was considered to be the main one [7].

Geological approaches tend to take into account the state of the earth and some indicators that are typically associated to earthquakes. One example is the temperature of the ground water resources, considered for the Tohoku earthquake [8]. This earthquake was also anticipated based on the geological data- mainly sandy deposits in coastal zones-that were subsequently flooded with the following tsunami after the earthquake [9]. The powerful characteristic of the geological methods is that they can be used for ancient earthquakes, based on the traces these earthquakes have left to the present day.

One such example is the Meio earthquake in 1498 [10], in which coastal geology was used to estimate the damages caused by the earthquake and its subsequent tsunami. However, geological approaches have also the drawback of being not reliable in many cases. The mentioned example in [9] is an attestation to this fact.

However, with the emergence of geodetic approaches and their rapid improvement in time, the prediction of earthquake entered a new phase. Fast (near real time), accurate, and reliable observations enabled geoscientist to predict the movement in earth's crust, even to the millimeter accuracy for the plate tectonics [11]. One such approach is based on the satellite data, such as the geodetic data derived from GPS. Using these data and taking into account the preseismic movements assumption for (large) earthquakes, the time of occurrence of these phenomena can be estimated more accurately than either the seismological or geological approaches. For instance, [12] reports that based on the high-resolution GPS time series before the Tohoku earthquake, the time of this earthquake could have been predicted up to 12 seconds accuracy. This very good result should motivate the geodetic community to use geodetic approach as their main tool to deal with these kinds of problems. There are many works on the prediction problem using time series, including [13], [14], [15], [16]. Note that the machine learning predictors are extrapolant, in contrast to the traditional approximants [17], [18], [19], [20], [21], [22], [23], [24], [25], which are interpolant.

The success of geodetic approaches has motivated us to propose a new algorithm to use geodetic data for the purpose of lend movement prediction. We use Lidar data. The data derived from these instruments are accurate, fast, reliable, and high-resolution. Hundreds of thousands of accurate geodetic data can be accessed in the form of point clouds. Hence, it is logical to be able to use these data for the purpose of earthquake prediction, using Machine Learning (ML) algorithms that enable us to accurately predict the next outcomes of data in the time sequence, even for the case of magnitude of earthquakes [26].

The following are the contributions we make in this letter
1. Presenting a novel approach to use Lidar geodetic data for the purpose of land movement prediction before an earthquake
2. Presenting a case study for the Ridgecrest, California 2019 earthquake

## II. THE ML SCHEME FOR EARTHQUAKE PREDICTION

The ML method that we present is a supervised algorithm. This scheme has different steps. These are based on the cartesian ellipsoidal location $(x, y, z)$, time $t$, and time increment, $dt$, of the acquired data. The scheme is based on the

M. Kiani is graduated from the School of Surveying and Geospatial data Engineering, University of Tehran, Enghelab Square, Iran (e-mail: mostafakiani@alumni.ut.ac.ir).

UTM coordinates of the data, $(x_{UTM}, y_{UTM})$, together with the geodetic height $h$, and the time of data acquisition. The concepts of ergodicity and splines play important roles in the algorithm. The steps of this scheme are as the following.

1. Changing UTM coordinates to the ellipsoidal ones, both Cartesian and curve-linear, $(\phi, \lambda)$
2. Computing the geoid height, $N$, and residual potential, $T$, from formulae in [27], to be able to use the local ergodicity condition for $T$ [28], after applying the so-called 13-point-difference-star ellipsoidal smoothers [17]
3. Passing $(x, y, z, t)$ and $T$ to a machine learning algorithm for the prediction of next values
4. Converting the predicted $T$ to a new $h$, and comparing the predicted $h$ with its previous value to determine the change
5. Decision on the possibility of earthquake or creep, considering the values of change in the land, and time of the earthquake, based on the previous step and a threshold of critical land change, $q$

The fifth step is of critical importance in determining the creep or earthquake. If the land change between two steps ($d$) does not reach a critical value ($q$), the land has a slow rate of change, meaning it has creep. However, earthquake is characterized by the sudden movement in the ground, thus $d$ must exceed $q$.

The diagram if Fig. 1 fully describes the scheme.

### III. EXPERIMENTS: THE ML SCHEME FOR THE RIDGECREST, CALIFORNIA 2019 EARTHQUAKE PREDICTION

The proposed ML scheme is tested for the Ridgecrest, California 2019 earthquake. The data are taken from [29] and [30]. The historical data (Fig. 2) are at least for 3 years before the earthquake, around 10/21/2016. The post-earthquake data (Fig. 3) are for 7/4/2019 and 7/5/2019. The data contain 58843443 measurements, for an area of 15 Km$^2$ extent.

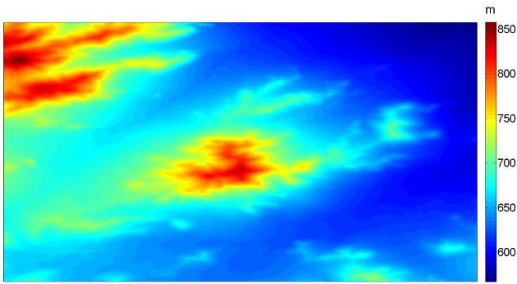

Fig. 2. Geodetic heights in the Ridgecrest region, before earthquake (2016)

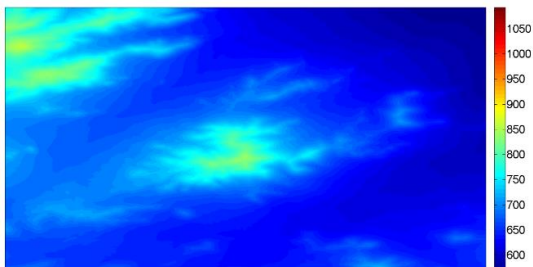

Fig. 3. Geodetic heights in the Ridgecrest region, after the earthquake (2019)

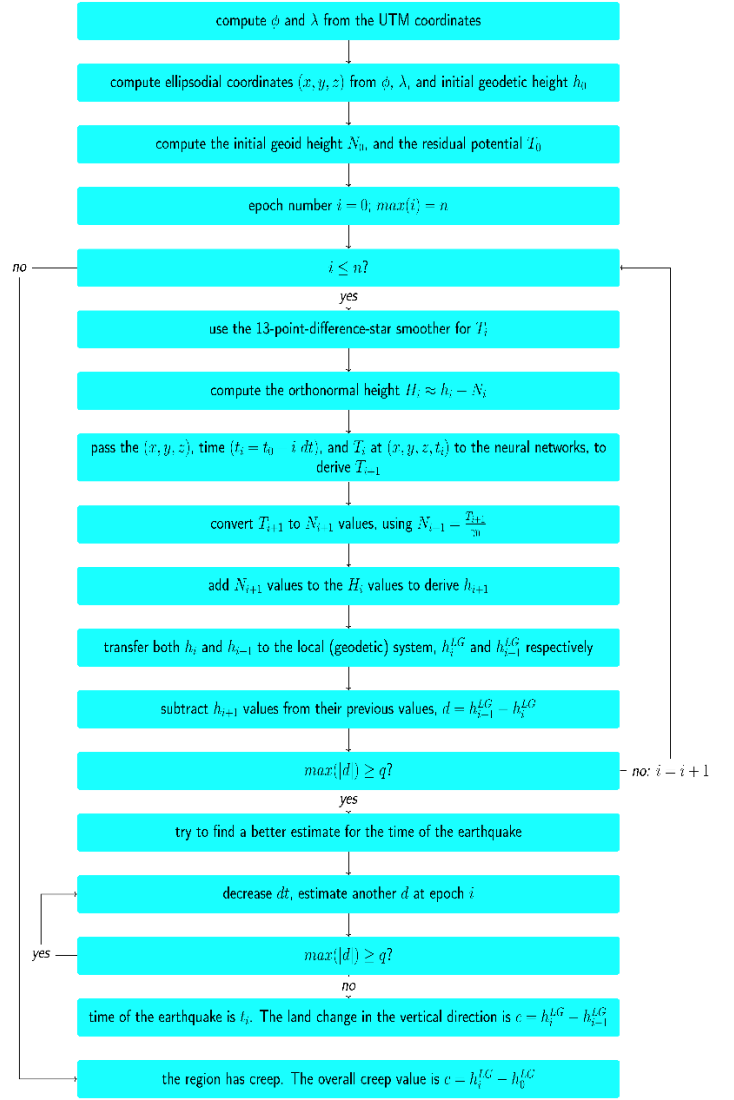

Fig. 1. Proposed algorithm for the earthquake or creep prediction

In order to implement the proposed algorithm, we set $q = 1m$, initial $dt = 14\ days$, and use the Multilayer Perceptron (MLP), Bayesian Neural Network (BNN), and Generalized Regression Neural Network (GRNN) [31] as the ML methods. It can be shown that an earthquake could have been predicted based on the algorithm.

After the algorithm in Fig. 1 is applied to these data, the time of earthquake and its land change are determined as the following, in TABLE I. Note that the Standard Deviation (StD) values are derived from the comparison between the predicted values at the exact time of the earthquake and their observed counterparts.

TABLE I
ACCURACY ASSESSMENT OF APPLYING THE ALGORITM IN FIG. 1 TO THE RIDGECREST DATA

| ML method | Time | StD(cm) |
| --- | --- | --- |
| MLP | 2 July | 14.22 |
| BNN | 2 July | 14.25 |
| GRNN | 2 July | 14.13 |

3As it can be understood from TABLE I, the GRNN method is the most accurate method. Its estimate of the time of the earthquake from its exact time differs almost 2 days.

## IV. Conclusion

In this letter, we proposed an algorithm by which the earthquake or creep prediction problem is addressed. A real study is presented for the Ridgecrest, California 2019 earthquake. It is shown this earthquake could have been predicted approximately 2 days before its happening. The most accurate ML method to use alongside the algorithm is GRNN. The promising results in this paper can be a point of departure for researchers studying in this important area of science.

Referencesbibliography[1] R. J. Geller, D. D. Jackson, Y. Y. Kagan, and F. Mulargia, "Earthquakes cannot be predicted," *science,* vol. 275, pp. 1616–1617, Mar. 1997.
[2] R. J. Geller, "Earthquake prediction: a critical review," *Geophys. J. Int.,* vol. 131, pp. 425-450, Apr. 1997.
[3] T. Rikitake, "Earthquake prediction," *Earth-Sei. Rev.*, vol. 4, pp. 245–282, Feb. 1968.
[4] S. Uyeda, T. Nagao, and M. Kamogawa, "Short-term earthquake prediction: Current status of seismo-electromagnetics," *Tectonophysics.*, vol. 470, pp. 205–213, Jul. 2009.
[5] R. J. Geller, "Shake-up time for Japanese seismology," *Nature,* vol. 472, pp. 408-409, Apr. 2011.
[6] C. R. Allen, "Responsibilities in earthquake prediction," *Bulletin of the Seismological Society of America*, vol. 66, no. 6, pp. 2069–2074, Dec. 1976.
[7] J. Liu, and Y. Zhou, "Predicting earthquakes: the Mw9.0 Tohoku earthquake and historical earthquakes in northeastern Japan," *Int. J. Disaster Risk Sci.*, vol. 3, no. 3, pp. 155-162, 2012.
[8] Y. Orihara, M. Kamogawa, and T. Nagao, "Preseismic changes of the level and temperature of confined groundwater related to the 2011 tohoku earthquake," *Nature Sci Rep.*, vol. 4, no. 6097, pp. 1–6, Nov. 2014.
[9] Y. Sawai, Y. Namegaya, Y. Okamura, and K. Satake, "Challenges of anticipating the 2011 earthquake and tsunami using coastal geology," *Geophys. Res. Let.*, vol. 39, no. L21309, pp. 1–6, Jan. 2012.
[10] O. Fujiwara, E. Ono, T. Yata, M. Umitsu, and Y. Sato, "Assessing the impact of 1498 Meio earthquake and tsunami along the Enshu-nada coast, central Japan using coastal geology," *Quaternary Int.*, vol. 308, pp. 4–12, Dec. 2013.
[11] C. Ito, H. Takahashi, and M. Ohzono, "Estimation of convergence boundary location and velocity between tectonic plates in northern Hokkaido inferred by GNSS velocity data," *Earth, Planets and Space,* vol. 71, no. 86, pp. 71-86, 2019.
[12] P. Psimoulis, M. Meindl, N. Houlie, and M. Rothacher, "Development of an algorithm for the detection of seismic events based on GPS records: case study Tohoku-Oki earthquake," 2013.
[13] M. Kiani, "On the suitability of generalized regression neural networks for GNSS position time series prediction for geodetic applications in geodesy and geophysics," *arXiv:2005.11106,* 2020.
[14] M. Kiani, "A precise machine learning aided algorithm for land subsidence or upheave prediction from GNSS time series," *arXiv:2006.03772,* 2020.
[15] M. Kiani, "Lateral land movement prediction from GNSS position time series in a machine learning aided algorithm," *arXiv:2006.07891,* 2020.
[16] M. Kiani, "A specifically designed machine learning algorithm for GNSS position time series prediction and its applications in outlier and anomaly detection and earthquake prediction," *arXiv:2006.09067,* 2020.
[17] M. Kiani, "Template-based smoothing functions for data smoothing in Goedesy," *Geodesy and Geodynamics,* to be published, doi: 10.1016/j.geog.2020.03.003, 2020.
[18] M. Kiani, N. Chegini, "Ellipsoidal spline functions for gravity data interpolation and smoothing," *Earth Observation and Geomatics Engineering,* vol. 3, issue 2, pp. 1-11, 2019.
[19] M. Kiani, "Comparison between compactly-supported spherical radial basis functions and interpolating moving least squares meshless interpolants for gravity data interpolation in geodesy and geophysics," *arXiv:2005.08207,* 2020.
[20] M. Kiani, "Local geoid height approximation and interpolation using moving least squares approach," *Geodesy and Geodynamics,* vol. 11, issue 2, pp. 120-126, 2020.
[21] M. Kiani, "Spherical approximating and interpolating moving least squares in geodesy and geophysics: a case study for deriving gravity acceleration at sea surface in the Persian Gulf," *Journal of Geodetic Science*, 2020.
[22] M. Kiani, "Image Gravimetry: A New Remote Sensing Approach for Gravity Analysis in Geophysics," *arXiv:2003.09388,* 2020.
[23] M. Kiani, "Optimal Image Smoothing and Its Applications in Anomaly Detection in Remote Sensing," *arXiv:2003.08210,* 2020.
[24] M. Kiani, "Identification and Classification of Phenomena in Multispectral Satellite Imagery Using a New Image Smoother Method and its Applications in Environmental Remote Sensing," *arXiv:2003.08209,* 2020.
[25] M. K. Shahvandi, "Numerical solution of ordinary differential equations in geodetic science using adaptive Gauss numerical integration method," *Acta Geodaetica et Geophysica,* vol. 55, pp. 277-300, 2020.
[26] A. Adeli, and A. Panakkat, "A probabilistic neural network for earthquake magnitude prediction," *Neural Networks*, vol. 22, pp. 1018–1024, May 2009.
[27] F. Barthelmes, "Definition of Functionals of the Geopotential and Their Calcu-lation from Spherical Harmonic Models," *Scientific Technical Report, GFZ German Research Center for Geosciences*, doi: 10.2312/GFZ.b103-09026, 2013.
[28] H. Moritz, "Advanced physical geodesy," Wichmann, California, 1980.
[29] M. J. Willis, W. D. Barnhart, R. Cassotto, J. Klassen, J. Corcoran, T. Host, B. Huberty, K. Pelletier, and J. F. Knight "CaliDEM: Ridgecrest, CA Region 2m Digital Surface Elevation Model," Funding by NSF and USGS, Data collection by Digital-Globe, Distributed by Open-Topography, doi:10.5069/G998854C, 2018.
[30] K. W. Hudnut, B. Brooks, K. Scharer, J. L. Hernandez, T. E. Dawson, M. E. Oskin, R. Arrowsmith, C. A. Goulet, K. Blake, M. L. Boggs, S. Bork, C. L. Glennie, J. C. Fernandez-Diaz, A. Singhania, D. Hauser, and S. Sorhus, "Airborne Lidar and Electro-Optical Imagery Along Surface Ruptures of the 2019 Ridgecrest Earthquake Sequence, Southern California," *Seismological Research Letters*, to be published, 2020.
[31] E. Alpaydin, "Introduction to machine learning," The MIT press, USA, 2014.